\documentclass[prl,superscriptaddress,floatfix,nofootinbib,twocolumn,aps]{revtex4-2}
\usepackage{graphicx,amssymb,amsmath,bm,dcolumn,xcolor}
\usepackage{threeparttable}

\usepackage[bottom]{footmisc}
\usepackage{comment}
\usepackage{xfrac}
\usepackage{appendix}
\usepackage{enumitem}
\usepackage[normalem]{ulem}
\usepackage{soul}
\usepackage{hyperref}
\usepackage{float}

\definecolor{dukeblue}{rgb}{0.0, 0.0, 0.65}

\begin{document}
\title{Searching for axion dark matter with magnetic resonance force microscopy}

\author{Elham Kashi}
\affiliation{Quantum Science and Engineering Program, University of Delaware, Newark, DE 19716, USA}

\author{Muhammad Hani Zaheer}
\affiliation{Department of Electrical and Computer Engineering, University of Delaware, Newark, DE 19716, USA}

\author{Ryan Petery}
\affiliation{Department of Electrical and Computer Engineering, University of Delaware, Newark, DE 19716, USA}

\author{Swati Singh}
\affiliation{Department of Electrical and Computer Engineering, University of Delaware, Newark, DE 19716, USA}
\affiliation{Department of Physics and Astronomy, University of Delaware, Newark, DE 19716, USA}
\email{swatis@udel.edu}

\date{\today}

\begin{abstract}
We propose a magnetic resonance force microscopy (MRFM) search for axion dark matter around 1 $\rm GHz$. The experiment leverages the axion's derivative coupling to electrons, which induces an effective A.C. magnetic field on a sample of electron spins polarized by a D.C. magnetic field and a micromagnet. A second pump field at a nearby frequency enhances the signal, with the detuning matched to the resonant frequency of a magnet-loaded mechanical oscillator. The resulting spin-dependent force is detected with high sensitivity via optical interferometry. Accounting for the relevant noise sources, we show that current technology can be used to put constraints competitive with those from laboratory experiments with just a minute of integration time. Furthermore, varying the pump field frequency and D.C. magnetic field allows one to scan the axion mass. Finally, we explore this setup's capability to put constraints on other dark matter - Standard Model couplings.
 
\end{abstract}

\maketitle

\textit{Introduction}.-- 
Dark matter is an integral part of the $\Lambda$CDM model of cosmology, but its fundamental nature remains a mystery \cite{bertone2018history}. There are several models of dark matter, with sub-eV mass bosons called ultralight dark matter (ULDM) being particularly compelling \cite{ferreira2021ultra,jackson2023search}. These consist of scalars (even-parity, spin 0), pseudoscalars (odd-parity, spin 0) and vectors (spin 1). Due to the large occupation number density required to constitute the observed local dark matter density, $\rho_{\rm DM}=0.4~\rm GeV/cm^3$ \cite{read2014local}, ULDM can be described as classical waves oscillating at Compton frequency $\omega_{\rm DM}=m_{\rm DM}c^2/\hbar$, where $m_{\rm DM}$ is the mass of the dark matter particle. The QCD axion is particularly well motivated, as it solves the strong CP problem of quantum chromodynamics \cite{peccei1977cp,peccei1977constraints,weinberg1978new,wilczek1978problem} and may be produced in large enough numbers in the early universe via the misalignment mechanism to account for all of dark matter \cite{preskill1983cosmology,abbott1983cosmological,dine1983not,ipser1983can}. In the post-inflationary PQ symmetry breaking scenario, the theoretically motivated mass range for QCD axions is $1~\rm \mu eV-100~\mu eV$ \cite{borsanyi2016calculation,dine2017axions,hiramatsu2011improved,kawasaki2015axion,berkowitz2015lattice,fleury2016axion,bonati2016axion,petreczky2016topological,ballesteros2017unifying,klaer2017dark,buschmann2020early,gorghetto2021more,buschmann2022dark}.

While dark matter is known to interact with baryonic matter gravitationally, ultra-weak non-gravitational couplings are present in most theoretical models, some of which may manifest as a real or effective A.C. magnetic field oscillating at $\omega_{\rm DM}$. For instance, the axion-photon coupling and the dark photon induce effective currents that generate real electromagnetic fields \cite{sikivie1983experimental, wilczek1987two, arza2022earth, chaudhuri2015radio}. Furthermore, axion-fermion couplings induce effective magnetic fields on fermion spins \cite{barbieri1989axion,alexander2018detecting}. Due to the wide viable mass range for ULDM, various technologies are required to fully search the allowed parameter space. This underscores a critical need for the development of ultra-precise A.C. magnetometers across a vast frequency range for dark matter detection.
Several dark matter searches are based on magnetic resonance techniques in magnetized samples \cite{crescini2018operation,crescini2020axion, flower2019broadening, ikeda2022axion,jackson2020overview, garcon2019constraints, walter2025search,xu2024constraining}.
Various other magnetometry techniques have been used to search for dark matter \cite{bloch2020axion, bloch2023constraints, lee2023laboratory, wei2025dark, bloch2022new, abel2023search, gao2022axion, karanth2023first, jiang2021search, gavilan2024searching}. 
In this manuscript, we propose a new mechanical, frequency-tunable search for dark matter using magnetic resonance force microscopy (MRFM).

\begin{figure}[H]
    \centering
    \includegraphics[width=\linewidth]{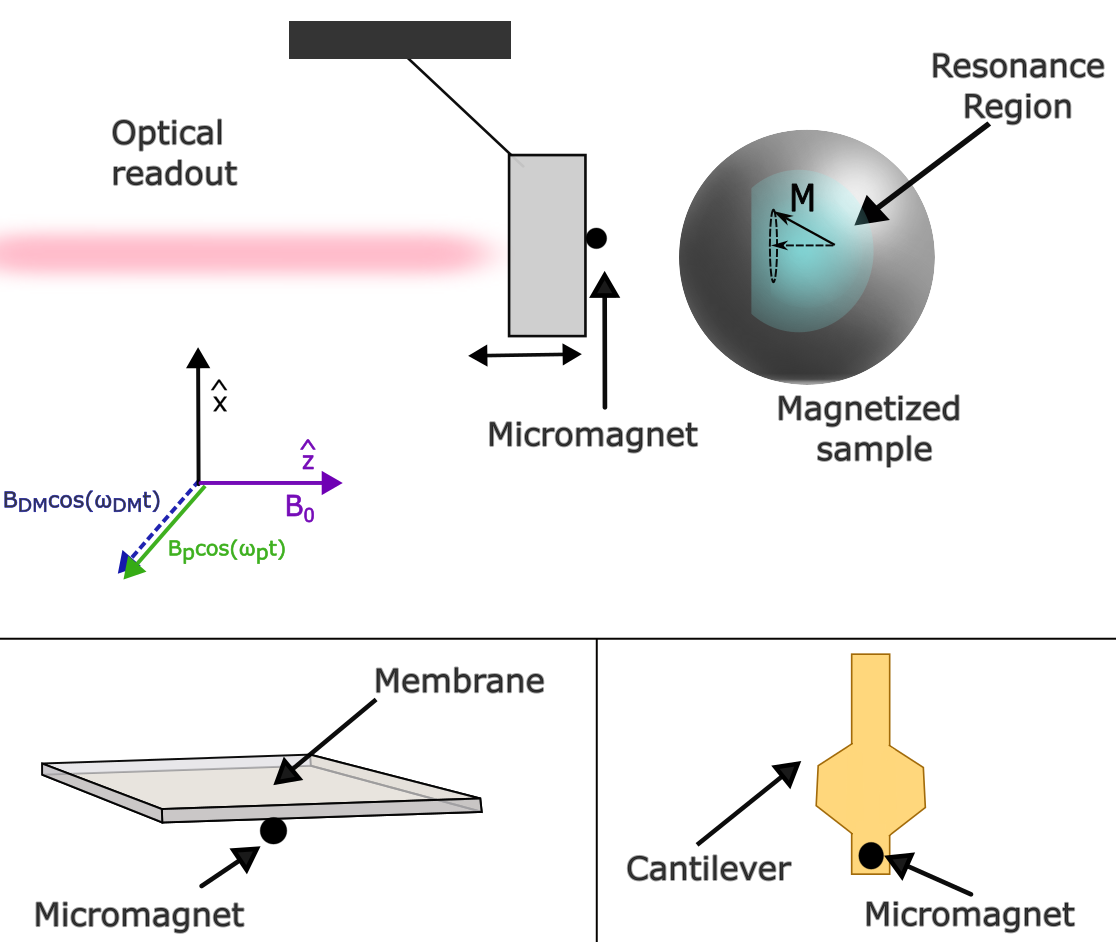}
    \caption{Diagram of the MRFM setup. The top illustrates the general scenario where a magnetized electron spin sample under influence of D.C. field $B_0\hat{z}$ and field of a micromagnet attached to a mechanical resonator undergoes magnetic resonance in the presence of a dark matter induced field $B_{\rm DM}\cos(\omega_{\rm DM} t)$. All spins within the resonance region put a force on the resonator. A pump field $B_{\rm p}\cos(\omega_{\rm p} t)$ parallel to the $B_{\rm DM}$ amplifies the dark matter signal and modulates it down to the micromagnet-loaded resonator's resonance frequency $ \vert\omega_{\rm DM}-\omega_{\rm p} \vert=\omega_{\rm m}$, inducing vibrations enabling optical detection. The bottom shows the two resonators considered in this work: membranes and cantilevers.} 
    \label{fig:Apparatus}
\end{figure}

We first recast several ULDM searches into an A.C. magnetic field detection problem. Then, we introduce a magnetic resonance force microscopy \cite{sidles1991noninductive, sidles1995magnetic, suter2004magnetic} detection scheme involving frequency down-conversion and optimize it for the axion-electron coupling. We show that the axion-electron coupling can be probed beyond current laboratory constraints with existing technology in under a minute of integration time. We then discuss the frequency scanning capabilities of our proposed setup. Finally, we explore this setup's capability to put constraints on other ULDM - Standard Model couplings.

\textit{Dark Matter Induced Magnetic Fields}.-- Here we describe various ULDM couplings to Standard Model fields that manifest as effective or real magnetic field effects.
Starting with axions, which are pseudoscalar particles that can be described by a classical wave of amplitude $\sqrt{2\rho_{\rm DM}}/\omega_{\rm DM}$ oscillating at its Compton frequency $\omega_{\rm DM}$. The axion field $a(\vec{x},t)$ can interact with fermions $\psi_f$ via the operator $g_{af}\partial_\mu a \bar{\psi}_f \gamma^\mu \gamma_5 \psi_f$, where $g_{af}$ is the axion-fermion coupling strength, $\gamma^\mu$ are the Dirac matrices and $f=e,p,n$ for electrons, protons and neutrons. This leads to an interaction Hamiltonian of the form $-\gamma_f \vec{S}\cdot \vec{B}_{\rm eff}$, where $\gamma_f=g_{f}e/2m_f$ is the fermion gyromagnetic ratio with the Lande g gactor $g_f$, $\vec{S}=\frac{\hbar}{2}\vec{\sigma}$ with Pauli matrices $\vec{\sigma}$, and $\vec{B}_{\rm eff}$ is an induced effective magnetic field \cite{barbieri1989axion,alexander2018detecting}. For axion couplings to electrons, protons and neutrons, respectively, the effective magnetic field strengths $B_{af}$ are given below in terms of the coupling strength $g_{af}$:
\begin{equation}
    B_{ae} \approx 2\times10^{-16}~\text{T} \times \frac{g_{ae}}{7\times 10^{-9}},
    \label{eq:B_ae}
\end{equation}
\begin{equation}
    B_{ap} \approx 5\times10^{-12}~\text{T} \times \frac{g_{ap}}{5\times 10^{-4}},
    \label{eq:B_ap}
\end{equation}
\begin{equation}
    B_{an} \approx 1\times10^{-11}~\text{T} \times \frac{g_{an}}{5\times 10^{-4}}.
    \label{eq:B_an}
\end{equation}
These equations give the magnetic field sensitivity required to probe the corresponding couplings to the level of current laboratory constraints in the $\rm GHz-THz$ range. 

Axions can also couple to photons and induce an effective current that generates real magnetic fields \cite{sikivie1983experimental, wilczek1987two, caputo2021dark, arza2022earth}. Here, a background magnetic field $B_0$ is required to mediate the interaction, and the induced magnetic field as a function of coupling strength $g_{a\gamma}$ is
\begin{equation}
    B_{a\gamma} = 3\times 10^{-20}~\text{T} \times \left(\frac{L}{10~\text{cm}}\right) \left(\frac{B_0}{25~\rm \mu T}\right) \frac{g_{a\gamma}}{10^{-9}~\text{GeV}^{-1}},
    \label{eq:B_agamma}
\end{equation}
where $L$ is the size of the experiment. 

In addition to axions, dark photons coupling to Standard Model photons can also induce an effective current that generates real electromagnetic fields \cite{caputo2021dark, arza2022earth}. The magnetic field generated by the dark photon as a function of the kinetic mixing parameter $\epsilon$ and experiment size $L$ is
\begin{align}
    B_{A^\prime} = 4\times 10^{-18}~\text{T} \times \left(\frac{f_{\rm DM}}{10^{9}~\rm Hz}\right) \left(\frac{L}{15~\rm cm}\right) \frac{\epsilon}{10^{-13}}.
    \label{eq:B_Aprime}
\end{align}
While better astrophysical constraints may exist for certain couplings, these may be more model dependent, necessitating new laboratory searches to supplement these constraints. 
See Appendix A for further details of the various dark matter induced magnetic fields. We now show how these weak signals can be detected by re-purposing an existing technology.

\textit{Magnetic Resonance Force Microscopy}.-- Magnetic Resonance Force Microscopy is a scanned-probe technique that combines imaging capabilities of magnetic resonance imaging with resolution of atomic force microscopy \cite{hammel2007magnetic}. Unlike conventional magnetic resonance detection, which inductively detects the electromagnetic signal of transverse magnetization, MRFM directly senses changes in longitudinal magnetization through a mechanical resonator \cite{sidles1993signal}.
MRFM experiments generally employ cyclical modulation of longitudinal magnetization $M_z$ at the mechanical resonance frequency, using techniques like amplitude modulation (AM) \cite{Suter2004, wago1997magnetic}. An advantage of this mechanical readout is avoiding circuit noise and associated backaction damping. Appendix B provides a more detailed comparison of MRFM with other magnetic resonance imaging techniques.  

A typical setup involves a micromagnet attached to the mechanical resonator placed directly above the sample, as shown in Fig.~\ref{fig:Apparatus}. 
For an extended sample that is placed sufficiently far from the micromagnet, the micromagnet can be treated as a point dipole \cite{suter2002probe, sidles91}, as detailed in Appendix B. The micromagnet produces a strong local magnetic field gradient that couples the resonator to magnetized spin sample underneath it. As a result, the resonator senses a force given by \cite{Suter2004},
\begin{align}
    F_z(t) = \int{-\left[\vec{m}(\vec{r},t)\cdot\vec{\nabla}\right]B_z(\vec{r}) d^3r},  \label{eq:generalFz}
\end{align}
with $\vec{m}(\vec{r},t)$ and $B_z(\vec{r})$ as magnetization density of the sample and the magnetic field of the micromagnet along the direction of mechanical motion, respectively. This integral is evaluated over the resonance area of the magnetized sample, as discussed below.

To induce magnetic spin resonance, a static magnetic field $B_0$ along the $z$ direction polarizes the sample, and an oscillating magnetic field $B_{\rm rf}$ is applied perpendicular to it. The bias field $B_0$ and the micromagnet's magnetic field $B_z$ determine the Larmor frequency $\omega_L=\gamma(B_0+B_z)$. Here, the role of the radio frequency field is played by the dark matter-induced magnetic field $B_{\rm DM} \cos(\omega_{\rm DM} t)$. To enhance the weak dark matter signal, a strong pump field $B_{\rm p} \cos(\omega_{\rm p} t)$ is applied parallel to $B_{\rm DM}$ with $B_{\rm p}\gg B_{\rm DM}$, in an approach similar to \cite{ruoso2016quax}.

The magnetization density $\vec{m}(\vec{r},t)$ of the magnetized sample can be calculated using stationary solutions of the Bloch equations \cite{abragam1961principles},
\begin{equation}
    \frac{d\vec{m}}{dt}= \gamma \vec{m} \times \vec{B} - \frac{{m_x \hat{x} +m_y \hat{y}}}{T_2} - \frac{m_z -m_0}{T_1}\hat{z},
\end{equation}
as long as $|\omega_{\rm DM}-\omega_{\rm p}|\ll\omega_L$ \cite{augustine2002transient, ruoso2016quax}. Here, $m_0 = n_s \mu_B \tanh(\mu_B B_0 / k_B T)$ is the equilibrium magnetization density, with $n_s$ as the spin number density, $T_1 $ and $T_2$ as the longitudinal and transverse relaxation times.
Due to the nonlinearity of the Bloch equations, the superposition of the two radio frequency fields, $ B_{\rm p} cos(\omega_{\rm p}) + B_{\rm DM} cos(\omega_{\rm DM})$, produces an on-resonance ($\gamma B_0-\omega_{\rm DM}=0$) beat-note modulation of $m_z$,
\begin{equation}
  \Delta m_z = m_z - m_0\simeq m_0 \gamma^2 T_1 T_2 B_{\rm p} B_{\rm DM} \cos(\omega_{\rm D} t), \label{eq:deltamz}
\end{equation}
where $\omega_{\rm D} =|\omega_{\rm p} - \omega_{\rm DM}|$ being the detection frequency satisfying the condition $\omega_{\rm D}^{-1} < \min \left\{ T_1, T_2 \right\}$. Eq. \ref{eq:deltamz} holds at resonance and for dark matter frequency around $\rm1~GHz$ $(\omega_L= \omega_{\rm DM} = \rm GHz)$. 
While a stronger pump field leads to larger amplification, the system must be kept far from saturation $B_{\rm p}\ll 1/\gamma\sqrt{T_1T_2}$. We choose $B_{\rm p} = 0.1/\gamma\sqrt{T_1T_2}$ in our calculations. For further details on the detection scheme, see Appendix B.

Choosing $\omega_{\rm D}=\omega_m$, this scenario becomes formally equivalent to applying an AM scheme in MRFM, with Eq.~\ref{eq:generalFz} giving the net force on the mechanical resonator,
\begin{equation}
F_z(t) = m_0 \gamma^2 T_1 T_2 B_{\rm p} B_{\mathrm{\rm DM}} \cos(\omega_{\rm D} t) \int \frac{\partial B_z(\vec{r})}{\partial z} d^3r .
\label{eq:net_force}
\end{equation}
The integral is performed over the resonant part of the sample volume, where the electron spins are in magnetic resonance as defined by the condition $\omega_{\rm DM} = \omega_{L}\pm\delta\omega$ with linewidth given by $\delta\omega = \frac{1}{T_2}$ far from saturation. Near the zero-probe-field resonance regime, the entire sample can be within the resonance region \cite{suter2004magnetic}, which gives a large force on the mechanical system, despite the weak field gradient.
To determine the sensitivity of a MRFM setup to radio frequency field $B_{\rm DM}$, we now consider the contribution of various noise sources to the resonator motion. 

\begin{figure*}
    \centering
    \includegraphics[width=\linewidth]{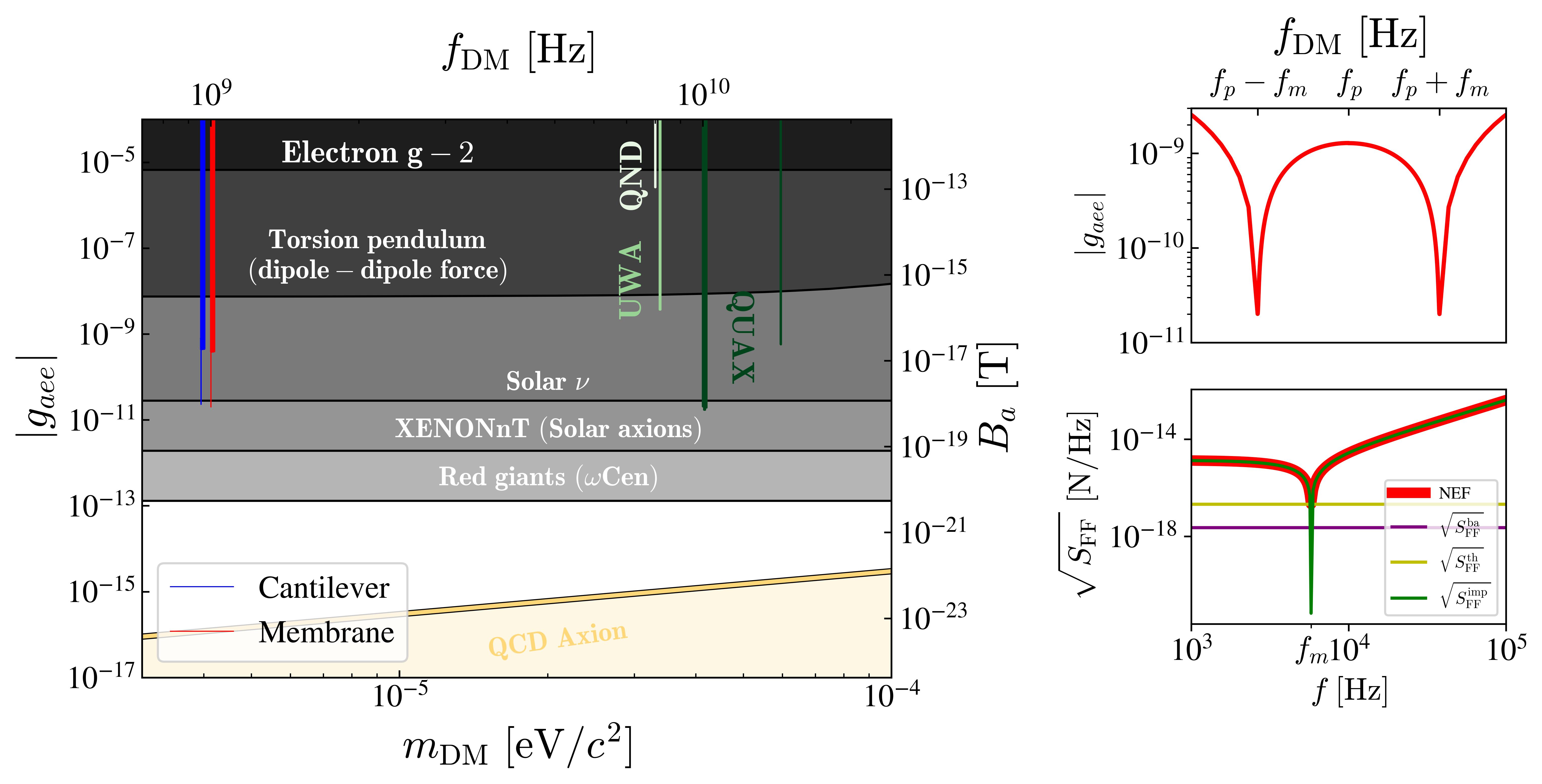}
    \caption{Sensitivity plot of axion-electron coupling strength $g_{aee}$ (left axis) and effective magnetic field (right axis) as a function of axion mass $m_{\rm DM}$ (lower axis) and corresponding Compton frequency $f_{\rm DM}$ (top axis). Our results for MRFM based axion search are given in red (membrane) and blue (cantilever). For each resonator, a 100 day run at a single frequency is shown and a 100 day scan of minute long runs at neighboring frequencies.
    The QCD axion band is shown in yellow and existing constraints from astrophysical \cite{capozzi2020axion}, space \cite{gondolo2009solar,aprile2022search} and laboratory experiments \cite{terrano2015short,yan2019constraining} are shown in gray. Also shown are narrowband constraints from other magnetic resonance based searches in green, labeled QUAX \cite{crescini2018operation, crescini2020axion}, UWA \cite{flower2019broadening} and QND \cite{ikeda2022axion}. The top right panel shows the finer details of a single sensitivity line including the two sidebands at $\omega_p\pm\omega_m$. The axion sensitivity region can be moved easily by varying the pump field frequency $\omega_p$. The bottom right panel shows the contributions from the back action noise (purple), thermal noise (yellow), and imprecision noise (green) to the total noise equivalent force (red).}
    \label{fig:axion-electron}
\end{figure*}

\begin{figure}
    \centering
    \includegraphics[width=\linewidth]{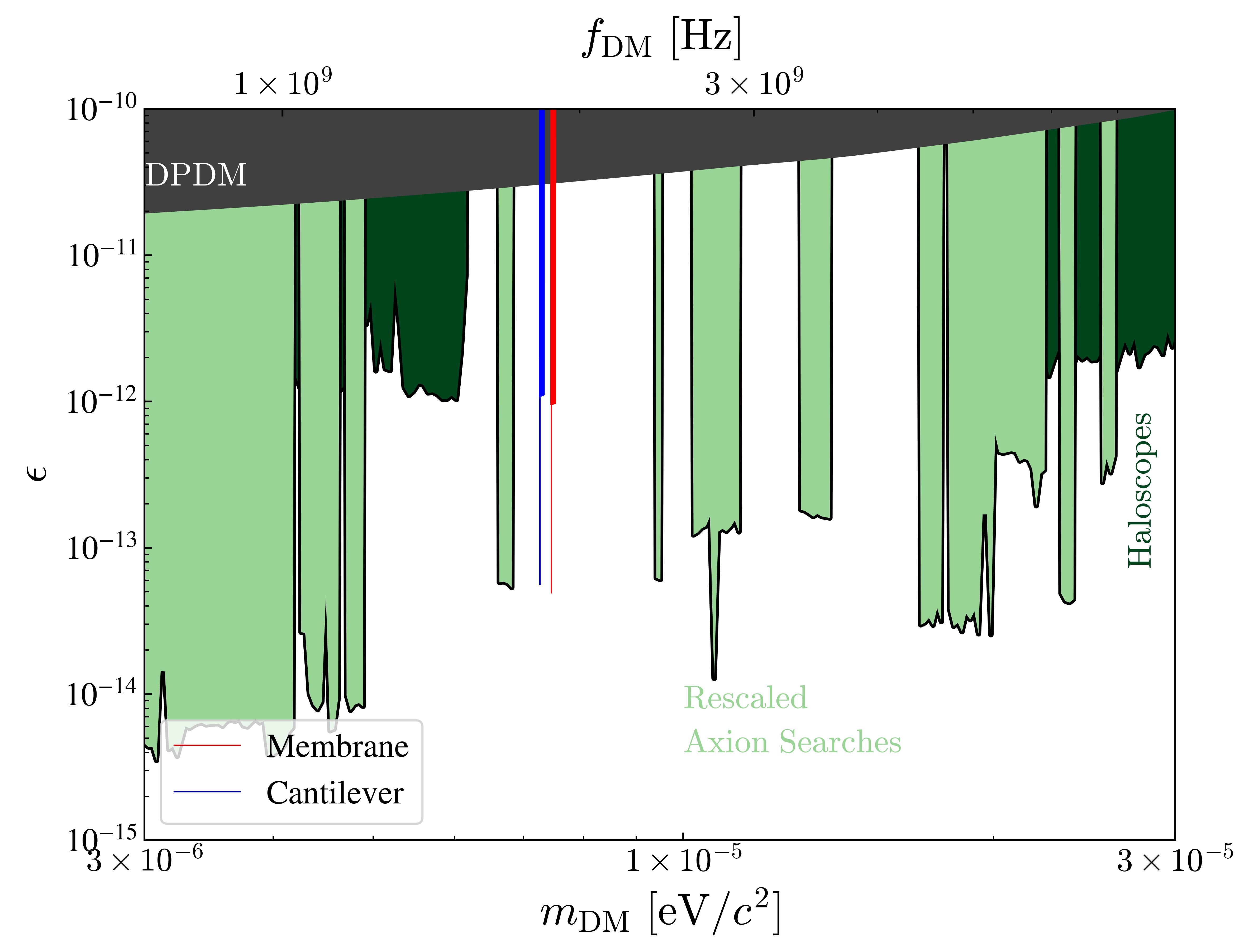}
    \caption{Sensitivity plot of dark photon kinetic coupling parameter $\epsilon$ as a function of dark photon mass $m_{\rm DM}$ (lower axis) and corresponding Compton frequency $f_{\rm DM}$ (upper axis). Our results for MRFM based axion search are given in red (membrane) and blue (cantilever). For each resonator, a 100 day run at a single frequency is shown and a 100 day scan of minute long runs at neighboring frequencies.
    The gray region represents constraints from astrophysical and cosmological observations assuming the dark photon makes up majority of dark matter, while dark green and light green are constraints from haloscope searches and rescaled limits from axion searches. See \cite{caputo2021dark} for details of various constraints on the dark photon kinetic coupling.}
    \label{fig:dark photon}
\end{figure}

\textit{Noise sources and minimum detectable coupling}.--
At mechanical resonance $\omega\sim\omega_m$, the sum of the measurement back action and imprecision noise is at a minimum, and the dominant noise source is due to the thermal bath of the mechanical system, with noise spectral density
\begin{equation}
    S_{FF}^{\rm th} = \frac{4k_B T m_{\rm eff} \omega_m}{Q},
    \label{eq:thermal_noise}
\end{equation}
where $T$, $Q$ and $m_{\rm eff}$ denote the temperature, quality factor and effective mass of the loaded mechanical oscillator, respectively.
After an integration time of $\tau_{\rm int}$, the minimum detectable force is then given by
\begin{equation}
    F_{\rm noise} (\omega)= \frac{1}{(t_{\rm int}\tau_{\rm c})^{1/4}} \sqrt{S_{FF}^{\rm NEF}(\omega)},
    \label{eq:F_noise}
\end{equation}
where the dark matter coherence time is $\tau_c \equiv \tau_{\rm DM} = \frac{c^2}{v_{\rm DM}^2 f_{\rm DM}} = \frac{10^6}{f_{\rm DM}}$ for couplings proportional to the field and $\tau_c\equiv\tau_{\nabla a}=0.68\ \tau_{\rm DM}$ for couplings proportional to the gradient of the axion field $a$ \cite{barbieri2017searching}. Here, NEF denotes the noise equivalent force, which corresponds to a signal to noise ratio of 1.

One can use Eq.~\ref{eq:F_noise} with Eq.~\ref{eq:thermal_noise} to calculate the thermal noise contribution to the force after integrating for $t_{\rm int}$. Assuming a signal-to-noise ratio of 1, we can then use Eq.~\ref{eq:net_force} to determine the minimum detectable magnetic field using our proposed setup,
\begin{equation}
    B_{\rm DM}^{\rm sen} (\omega_m) = \frac{\frac{1}{(t_{\rm int}\tau_{\rm c})^{1/4}} \sqrt{\frac{4k_B T m_{\rm eff} \omega_m}{Q}}}{m_0 \gamma^2 T_1 T_2 B_{\rm p} \int{\frac{\partial B_z(\vec{r})}{\partial z} d^3r }}.  
    \label{eq:B_DM_sen}
\end{equation}
The mechanical resonator has worse sensitivity away from the mechanical resonance due to large imprecision noise (see Appendix C for details). These expressions can be applied to any magnetically loaded mechanical system, provided that the underlying approximations hold.

Unlike several other magnetic mechanical systems, eddy-current effects do not pose a limitation here due to a combination of frequency mismatch and the material choice of YIG, which is not a conductor.

\textit{DM Parameter Space}.-- 
Before providing specific details of the device used for our estimates, we now overview existing constraints on the ULDM parameter space to work out the required magnetic field sensitivity. 
The parameter space for axion-electron coupling with existing laboratory and astrophysical constraints is shown in Fig. \ref{fig:axion-electron}. The most stringent constraint on the axion-electron coupling at $g_{ae}\sim10^{-13}~\rm eV$ stems from the lack of observed reduction in the tip of the red giant branch (TRGB) luminosity in globular clusters, where axion bremsstrahlung from electrons would otherwise cause significant energy loss and cooling \cite{capozzi2020axion}. Weaker constraints arise from solar axion searches, including limits on axion-induced solar energy losses from measurements of the neutrino flux \cite{gondolo2009solar} and null detection of electron recoils from solar neutrinos \cite{aprile2022search}. The best current laboratory constraint in the relevant mass range is due to torsion pendulum measurements of dipole-dipole forces between electrons \cite{terrano2015short} at $7\times10^{-9}~\rm eV$, while measurement of the anomalous magnetic moment of electrons \cite{yan2019constraining} also gives weaker constraints.

Three major milestones in the search for axions via their electron coupling are probing the coupling better than torsion pendulum limits \cite{terrano2015short}, observations of tip of the red giant branch (TRGB) luminosity in globular clusters \cite{capozzi2020axion}, and probing the QCD axion bands for DFSZ and KSVZ-like axions. These correspond to magnetic field strengths of $3\times10^{-16}~\rm T$, $5\times10^{-21}~\rm T$, and $2\times10^{-24}~\rm T$, respectively (see Fig.~\ref{fig:axion-electron}). The best current narrowband experimental constraints are from magnetic axion haloscopes that rely on detecting the axion-magnon coupling \cite{crescini2018operation, crescini2020axion, flower2019broadening, ikeda2022axion}. 
A number of proposals relying on optimistic advancements in technology have been proposed using QND measurements of axion-magnon conversion \cite{chigusa2020detecting, mitridate2020detectability, berlin2024physical, catinari2025hunting, chang2025mosaic}.
Our MRFM proposal for axion searches is similar in technical requirements and detection sensitivity to the best existing laboratory constraints, including QUAX \cite{crescini2018operation, crescini2020axion}. Unlike the aforementioned proposals, it does not require major technological advancements and the experiment can be performed with current technology. 

\textit{Device parameters.}-- 
As illustrated in Fig \ref{fig:Apparatus}, displacement of the micromagnet-loaded mechanical oscillator with frequency $\omega_m$ is detected optically via an interferometry setup. The dark matter frequency $\omega_{\rm DM}$ searched for by the experiment can be tuned by varying $\omega_{\rm p}$, the pump field frequency. Our scheme allows for a scanning search for axions using the same mechanical system, as mechanical resonance frequencies cannot be tuned significantly in most systems. 

Using the parameters given in Table \ref{Table:Parameters} and Eq. \ref{eq:B_DM_sen}, we find a minimum resonant magnetic field sensitivity at $\rm GHz$ frequency of about $1.6\times10^{-17}$ T ($8.4\times10^{-19}$ T) for a minute (100 days) of integration time with a membrane and $2.0\times10^{-17}$ T ($1.0\times10^{-18}$ T) for a minute (100 days) of integration time with a cantilever. Using Equations \ref{eq:B_ae} and \ref{eq:B_agamma}, we obtain minimum values that can be probed by our setup of the axion-electron coupling and the dark photon kinetic mixing parameters, respectively. These are plotted in Figs. \ref{fig:axion-electron} and \ref{fig:dark photon}. For the axion-electron coupling, an MRFM experiment can search for axions beyond existing laboratory constraints.
For the dark photon kinetic mixing, this system allows for filling in gaps between existing haloscope constraints.

\begin{table}[h!]
\begin{tabular}{|c|c|}
    \hline
    \multicolumn{2}{|c|}{\textbf{Parameters Used}} \\
    \hline \hline
    Temperature $T$ & $0.01~K$ \\
    Laser Power $P$ & $10^{-6}~\rm W$ \\
    \hline
    \multicolumn{2}{|c|}{Fabry-Perot Cavity} \\
    \hline
    Resonant Frequency $\omega_{\rm cav}$ & $1.77\times10^{15}~\rm rad/s$ \\
    Length $L_{\rm cav}$ & $10^{-4}~\rm m$ \\
    Finesse $F_{\rm cav}$ & $200$ \\
    \hline
    \multicolumn{2}{|c|}{Fe-Co Micromagnet}\\ 
    \hline
    Saturation Magnetization $\mu_0M_0$ & $2.4~\rm T$ \\
    Radius $R_{\rm mag}$ & $100~\rm \mu m$ \\
    Density $\rho_{\rm mag}$ & $8.5~\rm g/cm^3$ \\
    Mass $m_{\rm mag}$ & $36~\rm \mu g$ \\
    \hline
    \multicolumn{2}{|c|}{SiN Membrane} \\
    \hline
    Quality Factor $Q_{\rm m}$ & $10^6$ \\
    Resonant Frequency $\omega_{\rm m}$ & $2\pi\times5.8~\rm kHz$ \\
    \hline
    \multicolumn{2}{|c|}{Cantilever} \\
    \hline
    Quality Factor $Q_{\rm m}$ & $10^5$ \\
    Resonant Frequency $\omega_{\rm m}$ & $2\pi\times1.1~\rm kHz$ \\
    \hline
    \multicolumn{2}{|c|}{YIG Sample} \\
    \hline
    Longitudinal Relaxation Time $T_1$ & $10^{-6}~\rm s$ \\
    Transverse Relaxation Time $T_2$ & $10^{-6}~\rm s$ \\
    Number density $N_{\rm sample}$ & $2.1\times 10^{28}~\rm m^{-3}$ \\
    Length and Width & $2.5~\rm mm$ \\
    Height & $0.25~\rm mm$ \\
    \hline
\end{tabular}
\caption{Experimental parameters used in our calculations of the sensitivity of MRFM to dark matter-induced magnetic fields. The values for the resonators are chosen in the range of parameters from \cite{scozzaro2016magnetic,Takahashi2018,zwickl2012progress} for membrane and \cite{wago1998paramagnetic} for cantilever.}
\label{Table:Parameters}
\end{table}

The bottom right panel of Fig.~\ref{fig:axion-electron} shows contributions to the noise equivalent force spectral density from back action noise (purple), thermal noise (yellow) and imprecision noise (green). The top right panel shows the corresponding plot for sensitivity to the axion-electron coupling, showing zoomed in part of both sidebands $\omega_p\pm\omega_m$. This two-sideband feature comes from the frequency up conversion scheme and has been seen by, for example, the AURIGA detector \cite{branca2017search}. Our proposed experiment is limited by thermal noise over a relatively broad band with a linewidth of around $100~\rm Hz$ (see Fig. \ref{fig:Force_Noise} in Appendix C).

The target dark matter frequency $\omega_{\rm DM}$ can be scanned by varying the pump frequency $\omega_{\rm p}$ and the bias field $B_0$. Since the magnetic resonance linewidth ($\sim\rm MHz$) is much larger than the mechanical frequency, a significant scanning region can be probed with the same bias field value. With minute long runs, one can scan a $14.4~\rm MHz$ frequency range in $100$ days. 

One way to improve sensitivity is to increase the force on the resonator by using a larger sample. In free space, the sample size is primarily limited by radiation damping via magnetic dipole emission. A resonant cavity can be used to avoid radiation damping altogether \cite{crescini2018operation}, allowing for the use of large magnetized samples.


The constraints presented here involve just a minute of integration time on existing technology, with the entire experimental setup shown in figure \ref{fig:Apparatus} taking $\sim$1 cm$^3$ of volume. It is possible to fit multiple such setups inside a dilution refrigerator, improving the sensitivity by $\sqrt{N_e}$, for $N_e$ experimental setups, or scanning over a much larger bandwidth. A higher mechanical quality factor, as demonstrated on several platforms, can allow one to probe even weaker ULDM couplings. There is also potential for further multiplexing by using more magnetized samples, which only requires one to adjust the applied bias field to widen the resonance region. Furthermore, for axion-electron couplings in particular, flux concentrators can be used to amplify the signal \cite{gao2025focusing}. Finally, our proposal involves standard quantum limit readout, a limit that has already been circumvented in several experimental platforms, e.g. \cite{xia2023entanglement, jia2024squeezing, mason2019continuous}.

\textit{Conclusion and outlook}. --   
While most MRFM research has pushed towards detecting single spin flips, our work motivates exploration of the opposite regime, involving larger samples. Our approach involves a frequency scanning technique that can be adapted to other optomechanical searches for dark matter \cite{manley2020searching, manley2021searching, hirschel2024superfluid}. A similar MRFM based search can be conducted for axion-nucleon couplings using a nuclear spin sample, although the detection scheme may differ due to slower nuclear timescales. Our proposal for a magnetic resonance force microscope search for dark matter induced magnetic fields is just one thrust within the larger mechanical systems to search for fundamental physics theme \cite{carney2021mechanical, moore2021searching, kalia2024ultralight,ahrens2025levitated,bose2025massive,amaral2025first,qin2025mechanical,amaral2025towards,amaral2025morris}. Beyond dark matter, other interesting signals from gravitational waves, neutrinos, etc. can potentially be probed using optomechanical systems, paving the way for pushing forward technology and potentially discovering new physics.

\section{Acknowledgments}
 This work is supported by the National Science Foundation grant PHY-2047707 and the Office of the Under Secretary of Defense for Research and Engineering under Award No. FA9550-22-1-0323. We thank Kylee Smith for preparing Fig. \ref{fig:Apparatus}. We are also grateful to Yevgeny Stadnik and Jack Sankey for helpful discussions.

\appendix
\section{Appendix A: Dark Matter Couplings}	\label{DMproperties}

\textit{Axion-Fermion Coupling}.-- The axion field can be written as a classical wave in the form 
\begin{equation}
    a(\vec{x},t) = \frac{\sqrt{2\rho_{\rm DM}}}{\omega_{\rm a}} \cos(\omega_a t + \vec{k}_a\cdot\vec{x}+\dots),
\end{equation}
where the dark matter Compton frequency is related to its mass as $\omega_a=\frac{m_a c^2}{\hbar}$ and $|\vec{k}_a|=\frac{m_av_a}{\hbar}=\frac{10^{-3}m_a c}{\hbar}$.
Axions can couple to fermions via the operator
\begin{equation}
    \mathcal{L}_{af} = \frac{g_{af}\sqrt{\hbar^3 c}}{2m_f} \partial_\mu a \bar{\psi}_f \gamma^\mu \gamma_5 \psi_f.
\end{equation}
Here $f=e,p,n$ for electrons, protons and neutrons. We will ignore nuclear structure when dealing with nuclei. This leads to an interaction Hamiltonian analogous to that of a magnetic field interacting with the fermion spin
\begin{equation}
    H_{af} = -\gamma_f \vec{S}_f \cdot \vec{B}_{af},
\end{equation}
where the gyromagnetic ratio is $\gamma_f=\frac{g_f e}{2m_f}$, the spin operator is $\vec{S}_f=\frac{\hbar}{2}\vec{\sigma}_f$ and the effective magnetic field is 
\begin{equation}
    \vec{B}_{af} = \frac{2g_{af}\sqrt{\hbar c}}{g_f e} \frac{\vec{v}_{\rm DM}}{c^2}\sqrt{2\rho_{\rm DM}} \sin(\omega_a t),
\end{equation}
where the fermion g-factors are $g_e\approx 2, g_{\rm p}\approx5.58, g_n\approx-3.8$ and we ignore the spatial variation of the axion field.

\textit{Dark Photon Kinetic Mixing}.-- 
The dark photon ${A^\prime}^\mu$ can kinetically mix with with the Standard Model photon $A^\mu$, leading to the following interaction operator in the interaction basis \cite{fedderke2021earth, kalia2024ultralight}
\begin{equation}
    \mathcal{L}_{A^\prime} \supset \frac{\epsilon}{\mu_0} \left(\frac{m_{A^\prime}c}{\hbar}\right)^2 A_\mu {A^\prime}^\mu = - J_{\rm eff}^\mu A_\mu, 
\end{equation}
where $\epsilon$ is the kinetic mixing parameter. The interaction is equivalent to one between the EM field $A_\mu$ and an effective current $J^{\mu}_{\rm eff}$. 
Applying the Euler-Lagrange equations with respect to $A^\mu$ to $\mathcal{L}\supset-\frac{1}{4}F^{\mu\nu}F_{\mu\nu}-J^\mu_{\rm eff}A_\mu$, we obtain the equation of motion $\partial_\nu F^{\mu\nu}=J^\mu_{\rm eff}$. Using the equation of motion for massive vector fields $\partial_\mu {A^\prime}^\mu=0$, taking the non-relativistic limit, one can see ${A^\prime}^0=0$. Then, the dark photon field can be written as \cite{higgins2024maglev}
\begin{equation}
    |\vec{A}^\prime(\vec{x},t)|\approx \frac{\sqrt{2\rho_{\rm DM}}}{\omega_{A^\prime}}\cos(\omega_{A^\prime} t).
\end{equation}
Hence, the equation of motion $\partial_\nu F^{\mu\nu}=J^\mu_{\rm eff}$ gives the Ampere-Maxwell law
\begin{equation}
    \vec{\nabla}\times\vec{B} + \frac{1}{c^2}\frac{\partial\vec{E}}{\partial t}=\mu_0 \vec{J}_{\rm eff}.
\end{equation}
Therefore, the dark photon induced effective current generates real electromagnetic fields. When the dark photon Compton wavelength $\lambda_{A^\prime}=c/f_{A^\prime}=2\pi\hbar /m_{A^\prime} c$ is much larger than the experiment size $L$ (i.e. size of magnetic shield), the spatial variation of $\vec{J}_{\rm eff}$ can be ignored. In this regime, the magnetic field part dominates over the electric field term, since
\begin{equation}
    \left|\frac{\partial_t \vec{E}/c^2}{\vec{\nabla}\times\vec{B}}\right| \sim \frac{1}{c^2} \frac{\omega_{A^\prime}|\vec{E}|}{|\vec{B}|/L} = \frac{2\pi L}{\lambda_{A^\prime}} \ll 1.
\end{equation}
Hence, the primary effect of the dark photon is an oscillating magnetic field of amplitude
\begin{align}
    B_{A^\prime}  \sim \mu_0 J_{\rm eff} L \sim \frac{\sqrt{2\mu_0 \rho_{\rm DM}}c}{\hbar} \epsilon m_{A^\prime} L.
\end{align}

\textit{Axion-Photon Coupling}.-- 
Axions can also couple to photons via the following operator which can also be written as an interaction between the EM field and an effective current \cite{kalia2024ultralight}
\begin{equation}
    \mathcal{L}_{a\gamma} = \frac{g_{a\gamma}\sqrt{\hbar c^3}}{4\mu_0} a F_{\mu\nu} \tilde{F}^{\mu\nu} = J^\mu_{\rm eff} A_\mu,
\end{equation}
where $\tilde{F}^{\mu\nu}=\frac{1}{2}\epsilon^{\mu\nu\rho\sigma}F_{\rho\sigma}$ and the $J^\mu_{\rm eff}=-\frac{g_{a\gamma}\sqrt{\hbar c^3}}{\mu_0} \partial_\nu a \tilde{F}^{\mu\nu}$. Again, the zeroth component gives 0 and the current vector induces a magnetic field
\begin{equation}
    B_{a\gamma} =  \sqrt{2\hbar c\rho_{\rm DM}} g_{a\gamma} L B_0.
\end{equation}
Note that unlike the dark photon, the axion-photon induced magnetic field requires a background mediating magnetic field $B_0$.

\section{Appendix B: MRFM Details} \label{appendixB}

\textit{Comparing MRFM to magnetic resonance imaging.}-- Here we expand on some differences between magnetic resonance imaging and the MRFM scheme proposed here. 

We first expand on inductive vs. mechanical detection. In conventional inductive detection, the precession of a sample's magnetization induces a current in a pickup coil, which in turn generates a magnetic field that reacts back on the sample. This exchange of energy resulting in back-action noise, historically known as radiation damping, effectively damps the magnetization precession within the characteristic time $T_R = (2 \pi \eta  Q_{\rm coil} \gamma  \mu_0 M_0)^{-1}$, where $\eta$ is the filling factor (volume of sample per volume of coil ($V_s/V_c$)) and $Q_{\rm coil}$ is the coil's quality factor \cite{muller2007nuclear, ruoso2016quax}. MRFM overcomes this limitation by interacting with the sample through a magnetic force gradient (a one-way force detection) rather than a reciprocal electromagnetic induction (a two-way energy exchange). This allows for a less perturbative and therefore more sensitive measurement with a better signal-to-noise ratio \cite{sidles1993signal}.

While various modulation schemes enable detection of longitudinal magnetization, transverse magnetization detection in MRFM is challenging. This is due to the frequency mismatch between typical Larmor and mechanical frequencies. Even in the absence of this mismatch, spin control is complicated due to the requirement of a strong radio frequency field and non-applicability of the rotating wave approximation \cite{fischer2019spin}.

\textit{Micromagnet}.-- Here we expand on the theoretical treatment of the micromagnet. An integral part of an MRFM setup is the micromagnet that generates an inhomogeneous magnetic field with a spatial gradient that couples the mechanical resonator to the sample magnetic moments. In our setup, the micromagnet is approximated as a micro-sphere with radius $R_s$ and has saturation magnetization $M_0$, uniformly magnetized along $\hat{z}$. The criterion for the selection of the radius $R_s$ is its ratio with the distance between the center of the micromagnet and the surface of the sample ($d$), such that $R_s/d \ll 1$, and hence the magnet creates a dipolar magnetic field with strength $B(r)=\frac{4\pi}{3}M_{0}(R_{s}/r)^{3}$ at radial distance $r$ from its center \cite{suter2002probe, hammel2007magnetic, sidles91}. 
This gives the following field gradient
\begin{align}
\frac{\partial B(\vec{r})}{\partial z} &
=   \frac{-3 B(r)}{2r}(3\cos(2\theta)+1)\cos(\theta) 
\\ \nonumber
& + \frac{3 B(r)}{r} \sin(2\theta)\sin(\theta),
\end{align}
with the polar angle $\theta$ between the positive z-axis and the radial vector. 

For the material of the micromagnet, we considered Fe-Co because of its high saturation magnetization $\mu_0M_0 \sim 2.4~\rm T$ \cite{xue2011geometry, liu2008magnetic}.

\textit{Micromagnet Loaded Mechanical System}.-- Here we discuss how to calculate the resonance frequency of a micromagnet loaded mechanical resonator. For example, consider a $V_{m}=\rm 5~mm\times5~mm\times500~nm$ volume SiN membrane ($\rho_{m}=3.17~\rm g/cm^3$) with spring constant $k=60~\rm N/m$ loaded with a micromagnet of mass $m_{\rm magnet}=36~\rm\mu g$. Then the mechanical resonator frequency can be calculated using $\omega_m=\sqrt{k/m_{\rm eff}}$, where the effective mass is approximately $m_{\rm eff}\approx m_{\rm magnet}+\frac{m_{\rm mech}}{4}$. Using the above numbers, the magnet-loaded membrane frequency is $\omega_m\approx2\pi\times 5.8~\rm kHz$. For cantilever, we use the magnetically loaded resonance frequency $\omega_m=2\pi\times 1.1~\rm kHz$, which matches that of \cite{wago1998paramagnetic}. In general, the resonance frequency of the loaded mechanical system will be estimated numerically. 

\textit{Resonance Volume}.-- The micromagnet and the bias field $B_0$ define the region of space where magnetic resonance can be induced. For conditions $B_0 \ne B_{\rm rf}/\gamma$, the sensitive slice is a thin shell with spatial thickness $x_{sl} \approx \frac{\delta B}{\nabla B}$, where $\delta B \propto \frac{1}{T_2} $ is the intrinsic resonance linewidth of the sample. However, this concept breaks down when $  B_0 = B_{\rm rf}/\gamma$, where the influence of the gradient becomes small across most of the sample, hence known as zero-probe-field resonance. In this regime, the sensitive region is no longer a slice and expands to cover nearly the entire sample. Thus, each spin experiences a total field close to $B_0$, and this results in a strong signal despite the weak local field \cite{suter2002probe}. 
  
\textit{Changes in longitudinal magnetization}.-- 
The evolution of an ensemble of spins in a magnetized sample due to external magnetic fields (magnetizing field $B_0\hat{z}$, and a driving radio-frequency field $B_{\rm rf}$ in the x-y plane) is described by Bloch equations \cite{abragam1961principles}:
\begin{equation}
    \frac{d\vec{m}}{dt}= \gamma \vec{m} \times \vec{B} - \frac{{m_x \hat{x} +m_y \hat{y}}}{T_2} - \frac{m_z -m_0}{T_1}\hat{z},
\end{equation}
where $\vec{m}$ is the magnetization density vector, $\vec{m}=(m_x, m_y, m_z)$. 

For low d.c. fields, comparable in magnitude to the r.f. field, one instead considers the modified Bloch equations $\frac{d\vec{m}}{dt}= \gamma \vec{m} \times \vec{B} - \frac{m_z-\chi_0 (B_0+B_z)}{T_1}\hat{z} - \frac{m_x\hat{x}+m_y\hat{y}}{T_2}$ \cite{abragam1961principles}. The longitudinal steady-state solution is
\begin{equation}
    m_z = \frac{\chi_0}{\mu_0} B_0\big ( 1 + \frac{(\omega/B_0\gamma) (\gamma B_{\rm rf})^2 T_1T_2}{1+(T_2\Delta\omega)^2 + (\gamma B_{\rm rf} )^2T_1T_2}  \big ), \label{Blochmz} 
\end{equation}
where $\frac{\chi_0}{\mu_0} B_0 =m_0$, $B_0 \gamma =\omega_L$, $\Delta\omega = \omega_L -\omega_{\rm rf}$. At low driving field (as defined by $\gamma^2 B_{\rm rf}^2 T_1 T_2 \ll 1$), the spins excited at resonance ($\omega=\omega_L=\omega_{\rm rf}$) vary the longitudinal magnetization to $m_0 + \Delta m_z$, where $\Delta m_z = m_0 \gamma^2 B_{\rm rf}^2 T_1T_2$ \cite{wago1997magnetic}. 

In our scenario ($B_{\rm rf} =  B_{\rm p}+ B_{\rm DM}$), leading to
\begin{align}
      B_{\rm rf}^2&= \frac{B_{\rm p}^2}{2} [1+\cos(2\omega_{\rm p}t)] + \frac{B_{\rm DM}^2}{2}[1+\cos(2\omega_{\rm DM}t)] \nonumber \\ 
    &+ 2 B_{\rm p} B_{\rm DM} \frac{\cos((\omega_{\rm p}-\omega_{\rm DM})t)+\cos((\omega_{\rm p}+\omega_{\rm DM})t)}{2}. 
\end{align}
Here, The D.C. terms $B_{\rm p}^2$ and $B_{\rm DM}^2$ give constant shifts, which are not useful for detection. Meanwhile, the high-frequency terms $\cos(2\omega_{\rm p}t)$, $\cos(2\omega_{\rm DM}t)$ and $\cos((\omega_{\rm p}+\omega_{\rm DM})t)$, oscillating at frequencies of the order of GHz, are inaccessible to the mechanical resonator. Therefore, only the difference frequency term $\cos((\omega_{\rm p}-\omega_{\rm DM})t)$ contributes to the signal. Since $\omega_{\rm p} \approx \omega_{\rm DM}$ and $\vert \omega_{\rm p} - \omega_{\rm DM} \vert \ll \omega_L$, the difference $\omega_{\rm D}= \vert \omega_{\rm p} - \omega_{\rm DM} \vert$ is small and can be detected by a mechanical resonator if the $\omega_{\rm D}$ coincides with the mechanical resonance $\omega_m=\omega_{\rm D}$. The $\omega_{\rm D}$ frequency component of $\Delta m_z$ thus obtained is given in equation 7 of the main text.


\textit{Magnetic-dipole radiation damping.-} For a magnetic sample in free space with a Larmor frequency and large magnetization $M_0$, the dominant loss mechanism is the magnetic dipole emission. This phenomenon can be understood classically as lost power in the form of emitted photons driven by the magnetic dipole, and is characterized by a relaxation time $T_r = \frac{c^3}{\omega^3} \frac{6\pi}{\mu_0\gamma M_0 V_s}$ \cite{griffiths2023introduction}. This gives a restriction on how large of a sample volume $V_s$ one can use. Alternatively, strategies like embedding the sample in a resonant cavity (as done by QUAX) may be employed to avoid this loss mechanism \cite{crescini2019towards,ruoso2016quax, crescini2018operation}.

\textit{Eddy-current effects.-} A possible concern in MRFM experiments is the noise from eddy-current effects in either the sample or the micromagnet, especially in A.M. methods. In our proposed setup, however, this effect can be neglected for two reasons. Firstly, a YIG sample has low conductivity and, therefore, negligible eddy currents. Secondly, the applied pump field oscillates at $\rm GHz$ frequencies, but the mechanical resonator operates in the $\rm kHz$ band, creating a large mismatch between the frequency of the induced eddy fields by the pump and the frequency for force detection via the resonator. As a result, even with a stronger pump field, the fast oscillating eddy fields generated inside
the micromagnet remain outside the mechanical detection bandwidth. 

\begin{figure}
    \centering
    \includegraphics[width=\linewidth]{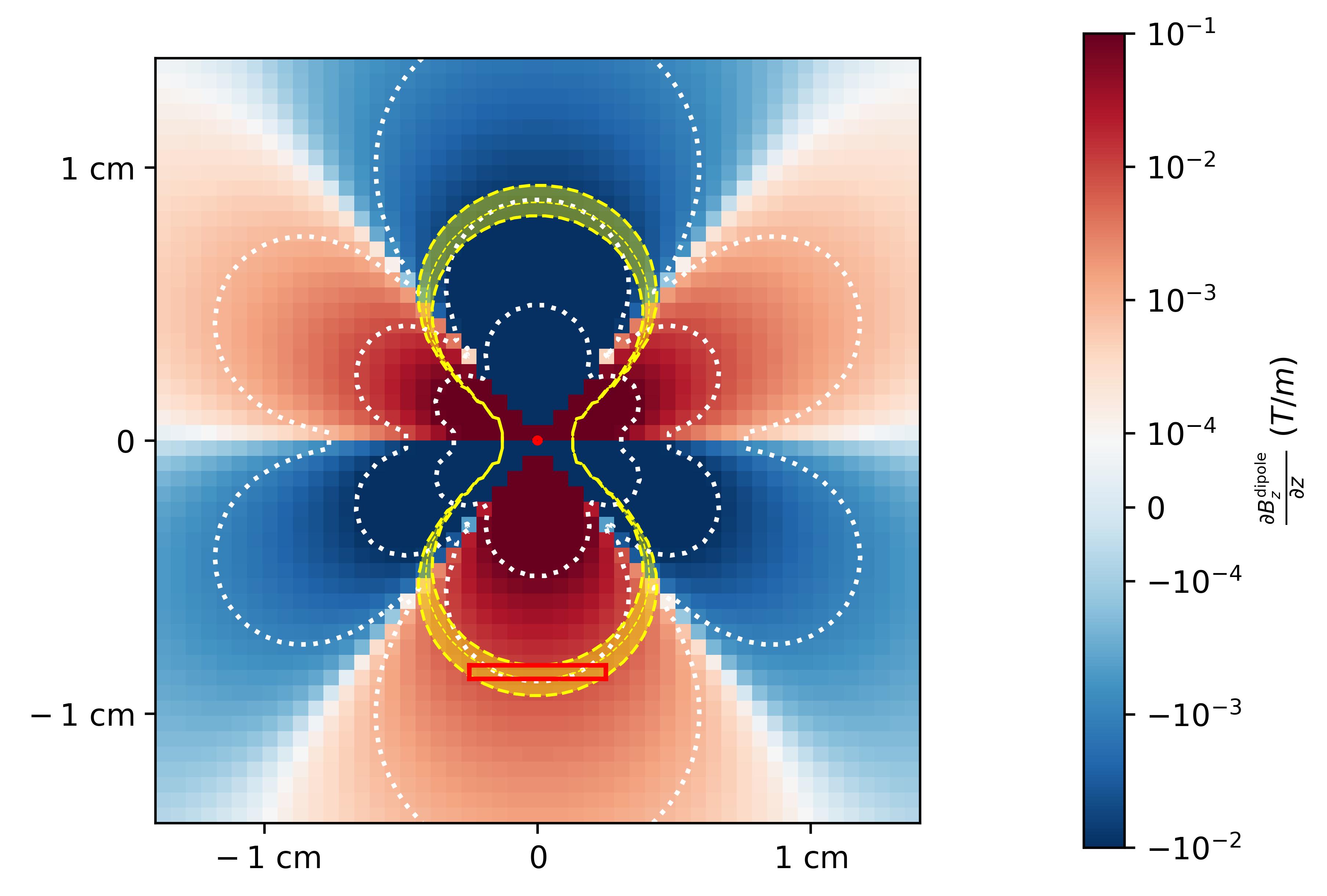}
    \caption{The magnetic field gradient produced by the resonator-attached micromagnet (shown as a red dot at the origin) is plotted as a contour against distance. The yellow region denotes the sensitive slice that is determined by the magnetic resonance condition $B_0+B_z = \frac{\omega_L(\vec{r})}{\gamma}$ and the linewidth $1/T_2$. 
    The red rectangle depicts the YIG sample.}
    \label{fig:resonant region}
\end{figure}

\section{Appendix C: Force Equivalent Noise}
\label{App:FEN}

The transfer function between force and position is obtained by solving the equations of motion for the system in the frequency domain \cite{aspelmeyer2014cavity}
\begin{equation}
    \chi_{xF}(\omega) = \frac{x(\omega)}{F_{\rm ext}} =\frac{1/m}{(\omega_m^2-\omega^2)+i\frac{\omega\omega_m}{Q}}.
\end{equation}
The oscillator response is given by
\begin{align} \nonumber
    S_{xx}(\omega) &= S_{xx}^{\rm imp}(\omega) + |\chi_{xF}(\omega)|^2 \times \\ & \left[S_{FF}^{\rm th}(\omega) + S_{FF}^{\rm ba}(\omega) + S_{FF}^{\rm \,signal}(\omega)\right].
\end{align}
The noises in the system are
\begin{align}
    & S^{\rm imp}_{xx} (\omega) = 2 \frac{\kappa}{16\bar{n}G^2}\left(1+\left(\frac{\omega}{\kappa/2}\right)^2\right), \\
    & S_{FF}^{\rm ba}(\omega) = 8\bar{n} \frac{4\hbar^2G^2}{\kappa}\frac{(\kappa/2)^2}{2} \frac{2}{(\kappa/2)^2+\omega^2},
\end{align}
and
\begin{align}
    S_{FF}^{\rm th} (\omega) = \frac{4k_BT}{\omega} \mathcal{I}m\left(\frac{1}{\chi_{xF}(\omega)}\right)
    \label{eq:fullThermNoise}
\end{align}

\cite{saulson1990thermal}, which are the imprecision, backaction, and thermal noise respectively. 

Here, $G=\omega_c/L$ is the optomechanical coupling \cite{aspelmeyer2014cavity} with cavity frequency $\omega_c$ and cantilever length $L$, $\kappa=\frac{\omega_{\rm FSR}}{F} = \frac{c}{2LF}$ with finesse $F$ is the cavity decay rate, and 
\begin{align}
    \bar{n} = \frac{\kappa P}{\hbar\omega_c (\kappa^2/4)},
\end{align}
is the mean occupation of the cavity, where $P$ is the laser power.

The imprecision noise is a shot noise from the discrete nature of photons, the backaction noise is due to the radiation pressure of the laser, and the thermal noise is due to thermal fluctuations of the system. While not obvious from equation \ref{eq:fullThermNoise}, the thermal noise in the system is white.

We now define a quantity known as the noise equivalent force (NEF). This is a force signal power such that it doubles the power at every frequency compared to the case of zero signal. In essence, this is a signal-to-noise-ratio (SNR) equal to 1 condition on the signal. It tells us how strong a possible signal would have to be at any frequency to be considered detectable. It can be found to be

\begin{equation}
    S_{FF}^{\rm NEF}(\omega) = |\chi_{xF}(\omega)|^{-2} S_{xx}^{\rm imp}(\omega) + S_{FF}^{\rm th}(\omega) + S_{FF}^{\rm ba}(\omega).
\end{equation}

The smallest force amplitude detectable in integration time $t_{\rm int}$ is given by 
\begin{align}
    F_{\min}(\omega) = \frac{1}{(t_{\rm int}\tau_{\rm c})^{1/4}} \sqrt{S_{FF}^{\rm NEF}(\omega)}.
\end{align}

The quarter-root scaling with integration time is due to the fact that the coherence time of the signal is much less than the integration time. Through a method of averaging the power spectral densities (PSDs) of several short-time measurements of length equal to 1 coherence time (known as Bartlett's method \cite{proakis2006}), we arrive at the quarter-root scaling \cite{manley2023searching}.

\begin{figure}
    \centering
    \includegraphics[width=\linewidth]{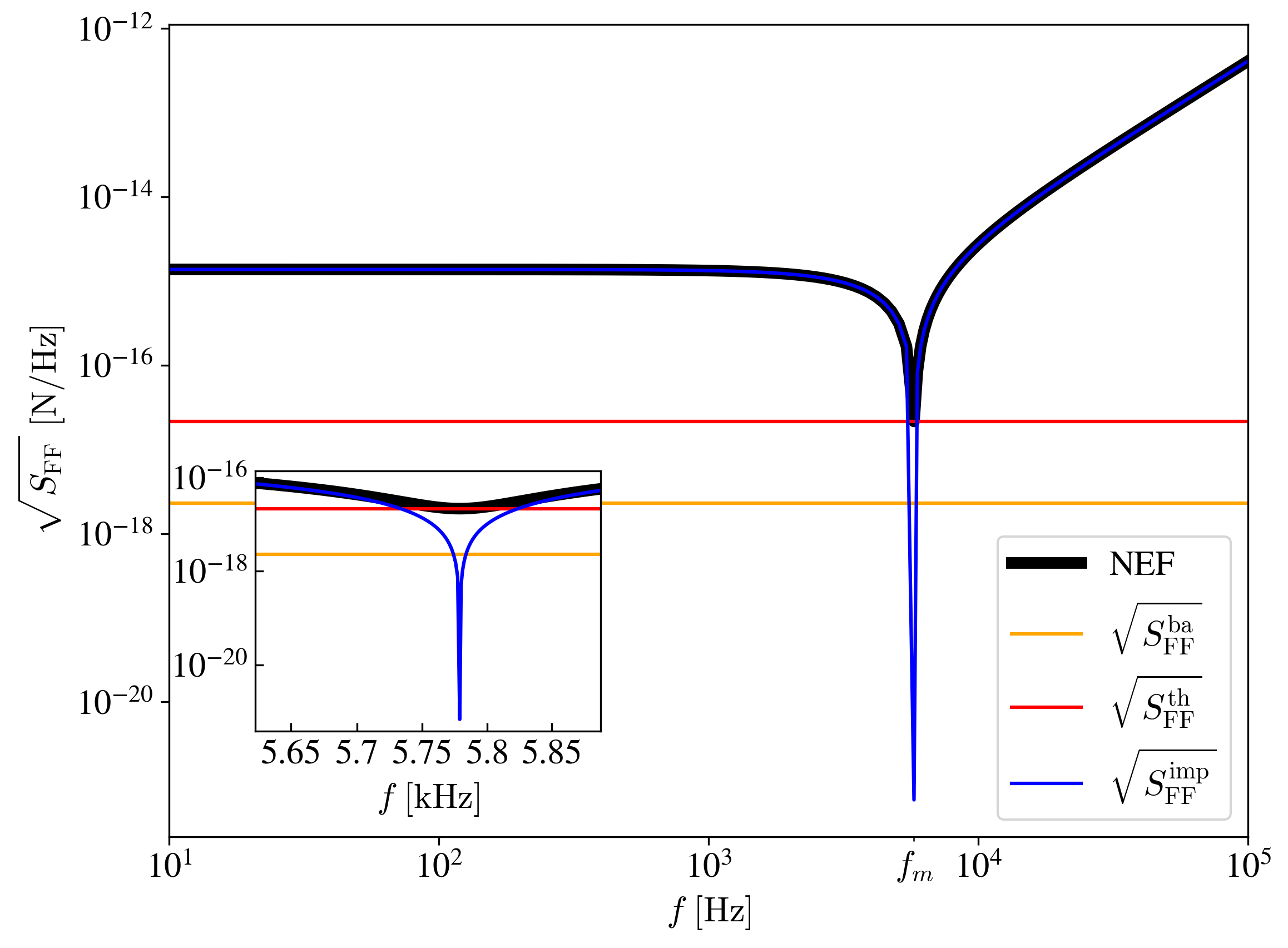}
    \caption{Noise contributions to the mechanical resonator motion due to back action noise (yellow), thermal noise (red), and imprecision noise (blue) are shown as the square root of the noise equivalent force spectral density as a function of frequency. The dominant part of the noise equivalent force (NEF), shown in black, is imprecision noise off-resonance and thermal noise on-resonance with the mechanical frequency $f_m$. The box on the lower left shows the zoomed in region around $f_m$. The bandwidth of the setup can be seen to be around $100~\rm Hz$.}
    \label{fig:Force_Noise}
\end{figure}

\bibliography{MRFMDMsensor}

\end{document}